\documentclass{article} 
\usepackage{colm2024_conference}

\usepackage{microtype}
\usepackage{url}
\usepackage{booktabs}
\usepackage{fontawesome}

\usepackage{url}
\usepackage{amsmath,amsfonts,bm}
\usepackage{listings}
\usepackage{graphicx}
\usepackage{amssymb}
\usepackage{multirow}
\usepackage{listings}
\usepackage{color}
\usepackage{makecell}
\usepackage{colortbl}
\usepackage{xcolor}
\usepackage{amsthm}
\usepackage{float}
\usepackage{wrapfig}
\usepackage{paralist}
\usepackage[figuresright]{rotating}
\usepackage{wrapfig,lipsum,booktabs}
\usepackage{subcaption}
\usepackage[labelformat=simple]{subcaption}
\usepackage{graphicx}

 \usepackage{algorithm}
\usepackage{algpseudocode}
\usepackage{algpascal}
\usepackage[misc]{ifsym} 

\usepackage[colorlinks=false]{hyperref}
\usepackage{threeparttable}

\title{Short-Term Effects of COVID-19 on Wages: Empirical Evidence and Underlying Mechanisms}  

\author{Bo Wu\thanks{Corresponding author. Email: \texttt{2021221257@stu.bisu.edu.cn}} \\
	\normalsize Beijing International Studies University \\
	\normalsize \faGithub \hspace{1mm}\url{https://github.com/sxjs1st}
}

\colmfinalcopy 
\begin{document}

\maketitle

\begin{abstract}
	This study investigates the causal relationship between the COVID-19 pandemic and wage levels, aiming to provide a quantified assessment of the impact. While no significant evidence is found for long-term effects, the analysis reveals a statistically significant positive influence on wages in the short term, particularly within a one-year horizon. Contrary to common expectations, the results suggest that COVID-19 may have led to short-run wage increases. Several potential mechanisms are proposed to explain this counterintuitive outcome. The findings remain robust when controlling for other macroeconomic indicators such as GDP, considered here as a proxy for aggregate demand. The paper also addresses issues of external validity in the concluding section.
\end{abstract}

\section{Background}
The COVID 19 pandemic has plagued world. Lockdowns, daily PCR tests, increasing unemployment and the new tendency of the remote working have reshaped the everyday life. For a long time, people suffered from loneliness. To make things worse, a lot of people were tortured by the economic recession caused by the pandemic due to the large-scale lockdown and its side effect, such as the breaking of the original supply chain\cite{almeida2021impact}. However, while recent related literature (Edesess and Loh,2023) or bigger reports like GLOBAL WAGE REPORT aim to provide a more comprehensive view on the COVID 19's impact on much larger markets like the whole labor market, few literature focuses on the specific impact on the wage. In that case, this situation arouses my curiosity to explore this field, finding out to what extent the pandemic affect people's wage provides. I believe I will discover an effective way to evaluate wage impact of the future epidemic, so that policy makers will be able to know the importance of appropriate policy corresponding to the disease to relieve people's burden.\cite{arndt2020covid}

The evidence is clear that the unemployment rate increased during the COVID 19. People left their position for multiple reasons and it was able to be anticipated that demand fell. Thus, it is normal to believe wage would fall. But the problem is that is just intuition inferred from previous recession. Is that real? Is there some special environment created by COVID 19 which changes the result?

\section{Aims}
This paper will explore how exactly the COVID 19 impacted the wage rate on a countrywide scale. I think the main advantage of my paper is that it utilizes both an Instrumental Variables method and an ordinary OLS to assess the potential bias and actively alleviating it.\cite{decker2020us} The main contribution is that it focuses on COVID 19's impact on the wage, effectively filling a gap in this field. 

I'll collect data from Johns Hopkins Coronavirus Resource Center to assess the seriousness of the COVID 19. They provide thorough time-series data for confirmed case number by country glob-ally. World population data from the World Bank are also collected to build my own per capita data set. For the assessment of the wage, I choose the data set from the GLOBAL WAGE REPORT published by International Labor Organization.\cite{douglas1923factors} Besides these data sets, I retrieve the Global Health Security (GHS) Index as well for the Instrumental Variable regression.

My regression result is not perfect, but it is enough for me to implement a short-term analysis and capture a short-term causal relationship. I find that in the short run, wage actually increased facing the COVID 19. Then on the topic of this phenomenon I propose some of the potential mechanism to explain the result. Then I conclude the whole paper and try to make some policy sugges-tions.

Apart from the current Section, the paper will be arranged as follows: Section is the literature review about the previous work related. Section introduces the model and fully explains the empirical method I used. Section describes the data sets with their sources and how I prepare and treat them to build my own adequate data sets for my study.
Section demonstrates and interprets the regression results, proposing potential mechanism with some further discussion. Section is the conclusion. At last there are appendices and references.

\section{Literature Review}
The wage has always been a central focus in the field of economic study, especially when facing the fallout of such a special period of COVID 19. The most comprehensive report on global wage may be the GLOBAL WAGE REPORT published by International Labor Organization. The report provides a concrete panel data set on the global wage. Recent newest GLOBAL WAGE REPORT (Geneva: International Labour Office, 2022) specially focuses on the COVID 19. The report offers thorough data and diagrams to show the decline of the wage on both world scale and regional scale. However, it lacks the analysis on the COVID 19 itself. The whole report largely focuses only on wage and gives on implications on the relationship between the seriousness of the COVID 19 and the wage. Also, the report generally provides more data than analysis on the wage.\cite{dui2022covid}

There are also some scholars trying to capture the wage trend under the pandemic on a regional level. Xavier and Mohit (2020) estimated the job loss and the wage decline in India for 2020. Anil(2020) basically did the same thing for Türkiye by constructing a possibility to work index.\cite{duman2020wage} Mayai, Augustino T. (2020) basically did the similar thing for South Sudan. However, these papers are essentially more like reports and lack real econometric analysis.

Mottaleb KA, Mainuddin M, Sonobe T (2020) explored the COVID 19's effect on income at an individual level based on the data from Bangladesh. It implemented econometric method but still lacks discussion on the COVID 19 seriousness. Lar-rimore, Mortenson and Splinter (2022) discussed the earning shocks in the US under the COVID 19, providing diagram showing a decline in annual earnings. Almeida, V., Barrios, S., Christl(2021) studyed the impact of the COVID 19 on household at EU-level using thethe EU microsim-ulation model, version I2.0+ and implies a regressive effect on household income.\cite{estupinan2020job} Arndt, Channing(2020) stressed the lockdown policy and other side effects' influence on income distribution in South Africa. These papers implemented a comprehensive econometric method while considering many macroeconomic variables. But still, they only explore at regional level, being unable to generate a more universal result. Also, they failed to using some measurement to quantify the seriousness of the COVID 19 in each countries in the world.\cite{gulyas2020consequences}

The gap in the topic of how wage reacts to the seriousness of the COVID 19 is enormous. So this paper aims to fill this gap by collecting data sets to evaluate related variables and conduct a regression analysis, trying to solve the problem on econometric level. Although GLOBAL WAGE REPORT lacks analysis on the COVID 19 part, the data attached in the GLOBAL WAGE REPORT are still useful for me to conduct my own analysis and I'd like to borrow that.
\section{Empirical Strategy}
\subsection{Model}

Since my aim is to find the causal relationship between the COVID-19 and the wage, the empirical model I choose is a regression, which can be represented as \textbf{Formula~\ref{e1}}.

\begin{align}
W_i = a + bE_i + u_i
\label{e1}
\end{align}

This is my main regression, where $W_i$ is the influenced wage, measured by the change in the real wage growth rate. $E_i$ is the independent variable, which is the COVID-19 case number accumulated in a year per capita. $u_i$ is the random disturbance. $a$ is the constant. Subscript $i$ indicates the year.

It can be anticipated that omitted variable bias will be large. Thus, to capture the causal relationship, the implementation of the instrumental variable is necessary, which can be represented as \textbf{Formula~\ref{e2}}:

\begin{align}
W_i = a + b(c + dZ_i) + u_i
\label{e2}
\end{align}

Simplify it, I get \textbf{Formula~\ref{e3}}:

\begin{align}
W_i = a + bc + bdZ_i + u_i
\label{e3}
\end{align}

In \textbf{Formula~\ref{e2}} and \textbf{Formula~\ref{e3}}, $Z_i$ is the instrumental variable. I'm going to let $Z_i$ be some medical level variable. $c$ is the random disturbance. $d$ is the measurement of medical level’s impact on the COVID-19 case number accumulated in one year per capita.

Since the coefficient $d$ is not relevant to my research question, I will not estimate the coefficient $d$. The coefficient $b$ directly reflects COVID 19's causal effect on the wage variable, which is our target coefficient. If it is positive, it implies that the COVID 19 made the real wage grow, vice versa. The magnitude of the $b$ coefficient can capture the sensitivity of the wage variable.

\section{Method}

As I have mentioned above, I mainly implement an instrumental variable method, i.e. a two-stage regression, to alleviate the omitted variable bias.The reason why the bias exists is that the impact of the COVID 19 is omnipotent. It not only affected the output and wage level, but also disrupted the whole economy. The chaos it caused might influence the wage level and be related to the COVID case number simultaneously. Issues like omitted variable bias and reverse causality are both possible to exist. I'll simultaneously conduct an ordinary OLS regression to indicate point.

I use the medical variable to mitigate all these biases. It is very unlikely to be systematically correlated with the change in real wage growth, but is sure to be directly correlated with the COVID 19 case number, rendering itself a perfect variable to perform an instrumental variable method. In this way, I can capture the casual relationship hidden behind all the biases.

\begin{table}[htbp]
\centering
\caption{\textbf{Summary statistics for the main variables in equation 3}}
\begin{tabular}{ccccccccc}
\toprule
 & \multicolumn{3}{c}{$E_i$} & \multicolumn{3}{c}{$W_i$} & \multicolumn{2}{c}{$Z_i$} \\
\cmidrule(lr){2-4} \cmidrule(lr){5-7} \cmidrule(lr){8-9}
Year & 2020 & 2021 & 2022 & 2020 & 2021 & 2022 & 2021 & 2020 \\
\midrule
Obs  & 94   & 94   & 94   & 94   & 84   & 38   & 94   & 94   \\
Mean & 2.306 & 2.173 & 13.335 & -- & 0.781 & -- & 48.741 & 48.707 \\
     &      &       &        & 2.660 &  & 2.624    &        &        \\
Mdn  & 1.823 & 4.252 & 8.466 & -- & -- & -- & 48.750 & 48.175 \\
     &      &       &        & 1.005 & 0.567 & 2.150 &       &       \\
SD   & 2.025 & 17.401 & 13.910 & 8.443 & 12.868 & 3.575 & 12.013 & 11.949 \\
Min  & 0.001 & -- & 0.003 & -- & -- & -- & 26.200 & 26.250 \\
     &      & 102.529 & & 45.932 & 38.503 & 10.200 &       &                \\
Max  & 7.765 & 19.258 & 55.167 & 23.434 & 95.545 & 6.800 & 75.900 & 76.050 \\
\bottomrule
\end{tabular}

\vspace{2mm}
\footnotesize{
\textbf{Note:} $E_i$ and $W_i$ are in percentage. $Z_i$ is an index. It is actually a score for each country, with 100 being the full score. All the data has been rounded to three decimal places using rounding rules.
}
\label{t1}
\end{table}
\begin{figure}[htpb]
	\centering
	\hspace{-1.5mm}
	\includegraphics[width=0.4\linewidth]{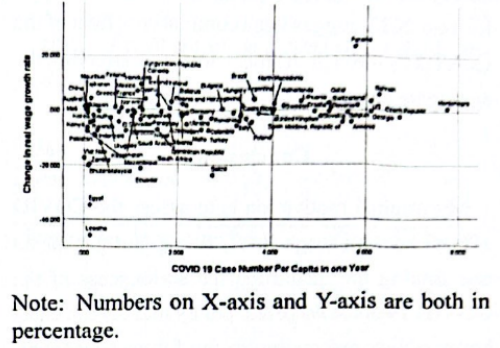}
	\caption{\textbf{Scatter plot for the Year 2020}}
	\vspace{-1.5mm}
	\label{f1}
\end{figure}


\begin{figure}[htpb]
	\centering
	\hspace{-1.5mm}
	\includegraphics[width=0.4\linewidth]{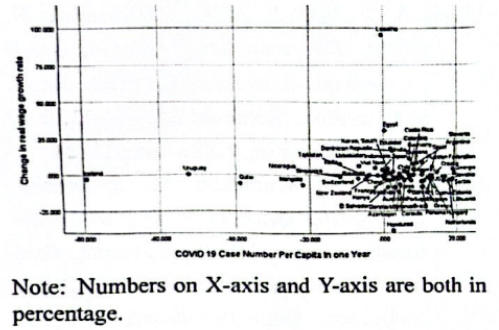}
	\caption{\textbf{Scatter plot for Year 2021}}
	\vspace{-1.5mm}
	\label{f2}
\end{figure}


\begin{figure}[htpb]
	\centering
	\hspace{-1.5mm}
	\includegraphics[width=0.4\linewidth]{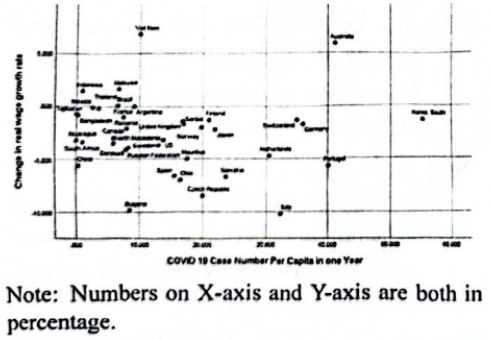}
	\caption{\textbf{Scatter plot for Year 2022}}
	\vspace{-1.5mm}
	\label{f3}
\end{figure}


\section{Data}
Since my main goal is to find the causal relationship between the impact of the COVID 19 and the wage level, I have to collect adequate data set to evaluate the both the COVID 19 seriousness and the wage change under the COVID 19.

The first data set I collected is the panel data from the Johns Hopkins Coronavirus Resource Center\footnote{\url{https://coronavirus.jhu.edu/}}. This data set provides global accumulated confirmed COVID-19 case number day by day, from January 22, 2020 to March 9, 2023 in the countrywide scale. Since my study is based on yearly scale, I don't need all that data. In this case, I just picked cross-sectional data of December 31, 2020, December 31, 2021 and December 31, 2022. Then, I subtracted the previous year's data from the current year, so I got the case number accumulated in one year, with exception of the Year 2020 due to being the first year of the pandemic.

Then, I collected yearly population data from the World Bank\footnote{\url{https://data.worldbank.org/indicator/SP.POP.TOTL}} for each country I selected. I implemented the treatment of division to calculate the yearly accumulated case number per capita. In this way, I built my own panel data set for independent variable to assess the seriousness of the COVID-19 in countries in the world within one specific year.

For dependent variable, it ought to be the assessment of the wage under the impact. To avoid issues like exchange rates and inflation, I chose the panel data set of the real wage growth by country from International Labor Organization’s \textit{Global Wage Report}\footnote{\url{https://www.ilo.org/digitalguides/en-gb/story/globalwagereport2022-23\#home}}. This data set provides real wage growth data by country from 2013 to 2022. To evaluate the impact of the COVID-19, I decided to use the change in real growth rate between years as dependent variable because I think it is able to subtract the original tendency from it, making this variable mainly reflects the impact of the COVID 19. It is a panel data set.

For instrumental variable $E_i$, what 1 chose is Global Health Security Index\footnote{\url{https://www.ghsindex.org/}} from Economist Im-pact, Nuclear Threat Initiative and Johns Hopkins Bloomberg School of Public Health, Center for Health Security. This index is a comprehensive panel data on national-level capacity across 195 countries to evaluate how well countries prevent, detect, and respond to pandemics. The Index not only provide overall index, but also offers sub-indexes about more specific field of health level. I only use the overall index. Since the data set downloaded lacks the data on Year 2020, I use the average of 2019 index data and 2021 index data to estimate the observations for the Year 2020. For Year 2022, I use the 2021 data too. The Global Health Security Index data are panel data.

Note that the data set on real wage growth from the GLOBAL WAGE REPORT is very incomplete. It lacks data on many small countries in the world, especially coming to the data of Year 2020 and following years, which makes it impossible for me to calculate the change in growth rate. Besides that, many data set like population and Global Health Security Index set also contains regional and territorial data like French Polynesia, which are not available for country-level data set on real wage growth. So I use the data set on real wage growth rate in Year 2020 as the base template. Any parts exceeding the template in other data sets are excluded manually by me, as shown in \textbf{Table~\ref{t1}}.

For $E_i$, the the standard deviation in 2020 is relatively low compared to other year's data. This may indicate the data from 2020 explain the relationship more significantly. The decrease in observation in Year 2021 and Year 2022 for the change in the real wage growth rate is clearly noticeable. This is a clear flaw in my data set, which may leads to statistical insignificance. However, I still decided to implement the regression because the scatter plots shows notable correlation or even potential causa-tion. The scatter plot \textbf{Figure~\ref{f1}} for 2020 can be referenced in this section. The scatter plots \textbf{Figure~\ref{f2}} and \textbf{Figure~\ref{f3}} for the remaining two years can be referenced below.
\section{Results}
\begin{table}[htbp]
\centering
\begin{threeparttable}
\caption{The effect of COVID-19 on wages in equation 3}
\begin{tabular}{lcccccc}
\toprule
\multicolumn{7}{c}{\textbf{Dependent Variable: Real Wage Growth Change}} \\
\midrule
\textbf{Year} & \multicolumn{2}{c}{2020} & \multicolumn{2}{c}{2021} & \multicolumn{2}{c}{2022} \\
             & (1) & (2) & (3) & (4) & (5) & (6) \\
             & OLS & IV  & OLS & IV  & OLS & IV \\
\midrule
Case Number in & 1.369\textsuperscript{***} & 5.146\textsuperscript{***} & 0.000 & -0.518 & -0.022 & 0.012 \\
one year per capita & (0.411) & (1.892) & (0.100) & (0.464) & (0.044) & (0.102) \\
\\[-1.5ex]
Constant & 5.817\textsuperscript{***} & 14.527\textsuperscript{***} & 0.780 & 2.462 & -2.283\textsuperscript{**} & -2.805 \\
         & (1.257) & (4.511) & (1.449) & (2.219) & (0.891) & (1.667) \\
\\[-1.5ex]
$R^2$    & 0.074 & 0.074 & 0.000 & 0.015 & 0.007 & 0.000 \\
Number of obs & 94 & 94 & 84 & 84 & 38 & 38 \\
\bottomrule
\end{tabular}
\label{t2}
\begin{tablenotes}
\small
\item \textit{Note}: For Year 2021 and 2022, the number of observations becomes smaller because of the lack of enough data on the change in real wage growth rate. There is standard error in the parentheses underneath the coefficients. * 10\%, ** 5\%, *** 1\% significance levels.
\end{tablenotes}
\end{threeparttable}
\end{table}
\subsection{Main Results}
My main goal is to estimate the impact of COVID 19 on wages, so I use regression and Instrumental Variable to evaluate the magnitude and the direction mainly by analyzing the coefficient of the main variable. The \textbf{Table~\ref{t2}} shows the regression results.

\textbf{Table~\ref{t2}} shows the regression results for each year. Since the pity is that only results for Year 2020 shows great statistical significance, my model can only answer part of my question, which means can only be restricted in 1-year short term.
Although I can only conduct my analysis in the short term of the year 2020, I think I can still derive interesting conclusions from the 2020 results and propose some potential hypothesis for the following year.

As the main variable coefficients (case number in one year per capita) show, the OLS coefficients greatly differ from the Instrumental Variable one, which is reasonable. Just like the year 2020, the ordinary OLS coefficient is upward biased. The reason for this is potential omitted variables. It is easy to anticipate that there are omitted variables which are negatively correlated with the change in real wage growth. For example, the lockdown policy caused by the COVID 19 disrupt the global value chain, causing chaos in the global economy. The implementation of the lockdown policy cannot be represented by the case number accumulated in one year per capita at all since only panic caused by the pandemic can urge the government to enact policy like social distance and lockdowns. Thus, the omitted variables greatly biased the coeflicient, proving my hypothesis in Method.

The impact of the COVID 19 on the change inreal wage growth is not a constant tendency as people expected to be a total negative impact. On the contrary, the data in 2020 shows a great positive effect on the change in real wage growth rate.One unit increase in the accumulated confirmed COVID 19 case in one year per capita in percentage is very likely to increase the real wage growth by 5.416 units in percentage in the short run (1-year period). Although the data from 2021 and 2022 do not show statistically significance, I can still observe a fluctuating result comparing the coefficient estimates. This can be seen from the \textbf{Figure~\ref{f2}} and \textbf{Figure~\ref{f3}}.

The unsolved problem is that the regression results for both 2021 and 2022 are not statistically significant at all. It may be an issue of the lack of the data for the following years. Another problem is that despite the insignificance, the coefficient estimate is fluctuating and the R-square for the fo)-lowing year is extremely small. In column (6) the R-square even rounds up to zero. All these indicators shows that my model either lacks enough data or does not fit the following two years at all. To solve these flaws, more data set may need to be gathered or more reliable model needs to be pro-posed.

Another problem about my model and results is that there may still be selection bias. As I have stated above, my observations are relatively small because many small and vulnerable countries in the world just failed to provide their statistics on the real wage growth to the International Labor Organization and my data set lacks them. It is obvious that larger countries with bigger economies are more likely to gather and maintain these statistical data facing the challenge of the COVID 19. Thosewhich are the weakest are usually unlikely to keep their statistical agency fully functional during pan-demic. That is to say, countries which performed better during the pandemic are more likely to appear in my observations, causing a selection bias, implying a potential upward biased result. But sadly I have no idea to alleviate it right now.

\subsection{Potential Mechanism}
I may try to propose a potential mechanism behind this short-term counter-intuitive phenomenon.

In the one-year short run, both aggregate supply and aggregate demand was disrupted by the COVID 19 and the following policies. Aggregate supply greatly declined during the first year of the COVID, mainly caused by the dismemberment of the global value chain due to the implementation of the global lockdown and social distancing policy.A large majority of the factories and business either stopped running or were forced to take a transition from the in-person working to the remote work-ing, effectively decreasing the efficiency. The aggregate demand should have been at least declined to the similar level in response to the impossibility of many entertainment service and other industries needed and used. However, things did not acted as what people expected. Many governments in the world actively intervened in the economy. \cite{larrimore2022earnings} For example, US enacted CARES Act on March 2020, signing 2.2 trillion dollars fiscal relief bill, including a sum of about 300 billion dollars direct payment to the individuals. Other stimuli to the aggregate demand include Paycheck Protection Program and emergency lendings to big financial institutions and banks. These measures successfully saved the aggregate demand from plummetlike what aggregate supply did and the people's expectation on the future. Supported by these aids, people were relatively unlikely to be eager to find jobs in the short run. This situation created a relatively tight labor market, because unlike other re-cession, laborers were stuck in home because of the COVID 19, rendering supply unable to increase in the short run. In this way. the global aggregate demand was relatively high compared to the aggregate demand as well, effectively increased the wage due to the scarcity of many products and ser-vices.\cite{loh2023covid} These factors collectively generates a positive coefficient and an increase related to the case number per capita in the short run. More cases per capita in a year means more seriousness of the COVID and the following impact on the aggregate supply in some specific country, thus rendering the wage increase more. This potential mechanism can clearly be seen in the scatter plot for the Year 2020, i.e. \textbf{Figure~\ref{f1}}, in which big economies like the US still maintain a positive change in the real wage growth while bearing a high case number per capita in a Year. On the contrary, small economies without enough power often failed facing the impact and suffered great negative change in the real wage growth rate.

In the long run, although I can derive no specific economic conclusions from the statistically insignificant results, I may still be able to propose a potential mechanism behind the phenomenon. In long-run period like Year 2021 and 2022, other cumulative factors may affect the wage. For example, in 2020 and early 2021, the US used to lower the interest rate to low levels to stimulate the economy.\cite{mayai2020economic} However,it is impossible to infinitely maintain a rather low rate. Once the interest cycle reachedthat specific time when the Fed was about to raise the rate, the aggregate demand is predictable to de-crease, disrupting the economy. Factors like this are not included or reflected in my model, which may be the main cause that my model does not fits the 2021 and 2022's data. I have tried to contain the GDP change rate as the control variable that represents the change in aggregate demand but the results are still insignificant statistically, which means further research is needed for the long run.

\begin{table}[htbp]
\centering
\begin{threeparttable}
\caption{Robustness check}
\begin{tabular}{lccc}
\toprule
\multicolumn{4}{c}{\textbf{Dependent Variable: Real Wage Growth Change}} \\
\midrule
\textbf{Year} & 2020 & 2021 & 2022 \\
\midrule
Case Number in one year per capita & 1.413\textsuperscript{***} & 0.017 & -0.038 \\
 & (0.432) & (0.100) & (0.040) \\
\\[-1.5ex]
Change in GDP growth rate & 0.070 & -0.295 & 0.510\textsuperscript{***} \\
 & (0.201) & (0.236) & (0.166) \\
\\[-1.5ex]
Constant & -5.402\textsuperscript{***} & 3.919 & -1.144 \\
 & (1.738) & (2.903) & (0.883) \\
\\[-1.5ex]
$R^2$ & 0.109 & 0.019 & 0.218 \\
Number of obs & 94 & 84 & 38 \\
\bottomrule
\end{tabular}
\label{t3}
\begin{tablenotes}
\small
\item \textit{Note}: For Year 2021 and 2022, the number of observations becomes smaller because of the lack of enough data on the change in real wage growth rate. There is standard error in the parentheses underneath the coefficients. * 10\%, ** 5\%, *** 1\% significance levels.
\end{tablenotes}
\end{threeparttable}
\end{table}
\subsection{Robustness Checks/Alternative Models}
To check robustness and explore if there is other better models for my study, I run a multivariate regression to see if adding a new variable alters my result or institutes a better model.\textbf{Table~\ref{t3}} I choose the GDP change rate in one year to take the aggregate demand into consideration. The results generates a similar conclusion: while the coefficient of the COVID 19 case number per capita in a year truly fluctuate with the introduction of the new variable, it still generate a statistically positive result in the short run, suggesting the robustness. It is worthing noting that GDP variable is statistically significant for year 2022, suggesting a cumulative effect of the COVID 19 through the mechanism of affecting the aggregate demand.

\section{Conclusion}
My original motivation is to assess the COVID
19's impact on wage level, filling the gap of no one linking the quantified the seriousness of the COVID 19 to the wage and policy makers can enact better policy responding to the future pandemics.This paper uses population data from the World Bank, COVID 19 confirmed case panel data from Johns Hopkins Coronavirus Resource Center and
GHS Index to construct the author's own data set and implement both an instrument variable regression and an ordinary regression for 3 years after the breakout of the COVID 19. The result from the regression enables me to conduct a short-term analysis which states that in the short run.\cite{mottaleb2020covid} My paper suggests that the COVID 19 actually has a positive effect of 5.146 in the change in the real wage growth rate per confirmed case number per capita in a year in the short run. And then, I present several potential mechanisms behind this phenomenon. My suggestion is that don't be deceived by the wage growth during the early days of the pandemic because my paper suggests that it can be just in the short run! For limitations, this paper can only answer part of my research question because it shows no significant result for the long run.

Since my study is on the global scale and utilize the countrywide data, my results and coefficient estimate should be applicable to most countries in the world in the short run when conducting short term analysis. However, there do exist some potentiallimitations. As I have mentioned above, my data set lacks data for many small countries and island countries in the ocean in the world. Thus, it is possible that my model is unable to be implemented to small countries even in the short run due to the lack of their observations in the regression and the actual impact on these small countries is negative.\cite{wang2022estimating} Mechanically speaking, the reason for this could be these small countries cannot face the COVID like a regular country at all. Maybe the COVID 19 just disrupted everything so they were just unable to take actions to relieve the crisis brought by COVID 19. Maybe these economies were so small that they could only rely on other countries, and other countries implemented policies that had negative effect on them. But for most countries, it should be true that in short run COVID 19 raised the wage for workers.\cite{ILO_Global_Wage_Report_2022}

As my potential mechanism part stated above. the positive effect generated by the COVID 19 may be a result of relatively high aggregate demand. In that case, I recommend future scholars should find more data and construct more advanced econometric model to find if that is true. Variables like GDP representing the aggregate demand should be taken into consideration in the new model. \cite{weeraratne2023covid}My view is that aggregate demand may not be a linear relationship with the wage level because the result of one of my multivariate linear regression model containing GDP by country as a variable shows no statistically significance and a bad R-square, meaning the linear model does not fit at all. Also, policy response is another factor that is worth studying. Responsive Policy to the COVID 19 may play a major role in determination of the wage level. Constructing a policy index or policy dummy may be necessaryin the future research. And for the selection bias mentioned in the Section, maybe we really have to wait for years to wait the small countries to fully recover from the fallout of the COVID 19 and update the data to eliminates the selection bias thoroughly.

\bibliography{colm2024_conference}
\bibliographystyle{colm2024_conference}


\end{document}